# EXPLORING SPACE WEATHER FROM YOUNG SOLAR-LIKE STARS AS WINDOWS TO EXOPLANETARY HABITABILITY

*Submitted in response to STScI Call: "Building a Roadmap for Hubble Science into the 2030s"*

V. S. Airapetian[1], K. Namekata[1,2], K. France[3], T. Sextro[4], M. Jin[5], J. Hu[6], T. Shi[7], K. V. Getman[4], E. D. Feigelson[4], J. Schlieder[1], M. McElwain[1], K. G. Carpenter[1], D. Sur[1,8]

[1]*NASA GSFC, Greenbelt, MD;* [2]*Kyoto University, Kyoto, Japan;* [3]*University of Colorado, Boulder, CO;* [4]*Pennsylvania State University, University Park, PA;* [5]*LMSAL, Palo Alto, CA,* [6] *University of Alabama, Huntsville, AL,* [7]*SETI, CA,* [8]*The Catholic University of America, Washington, DC*

**Abstract.** Young solar-like stars are efficient generators of magnetic activity, superflares, coronal mass ejections (CMEs), and stellar energetic particles (StEPs). These phenomena drive the early evolution of stars and shape the habitability of exoplanets. The Hubble Space Telescope (HST), with its unmatched far-ultraviolet (FUV) and near-ultraviolet (NUV) sensitivity, provides a uniquely powerful window into these processes—one that no current or near-future facility can replicate. This white paper articulates four interconnected science questions that require Hubble's continued operation and targeted observing programs over the next 10–15 years, enriched by new multi-wavelength insights from deep X-ray surveys of open clusters. We describe required instrument capabilities, critical synergies with contemporaneous missions (JWST, Chandra, XMM-Newton, TESS, and the Nancy Grace Roman Space Telescope (RST)), and the fundamental role Hubble observations will play in calibrating and informing the design of the Habitable Worlds Observatory (HWO). We advocate for large-scale coordinated campaigns targeting young solar-like stars as the highest-priority science program for the coming decade.

## 1. Introduction: The Far Ultraviolet Lives of Young Stars

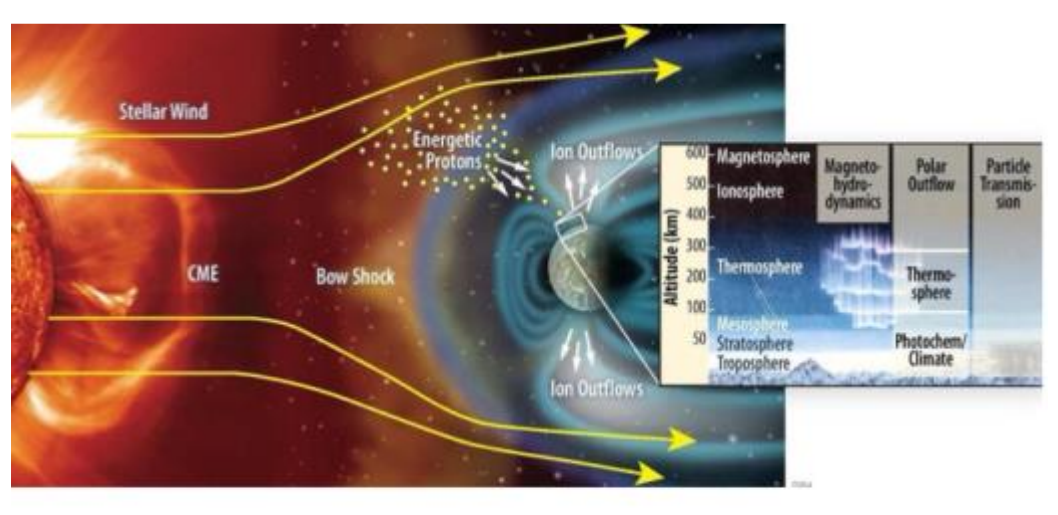

Fig.1. Schematic view of the complex exoplanetary space weather system that incorporates the physical processes driving stellar activity and associated SW including stellar flares, CMEs and their interactions with an exoplanetary atmosphere driven by its internal dynamics [1].

The Sun we observe today is a middle-aged, relatively quiescent star. Yet for the first billion years of its life, it was a far more magnetically violent object—spinning rapidly, erupting frequently, and bathing the nascent solar system in intense radiation and particle streams. Understanding this early epoch is not merely a historical curiosity; it is fundamental to answering whether the planets of young stars can ever become habitable, and how Earth itself came to support life. Young solar-like stars (spectral types F–K, ages ≤ 1 Gyr) are characterized by rapid rotation and deep convective envelopes, the prerequisites for efficient magnetic dynamos. The resulting strong, complex surface magnetic fields power X-ray bright coronae, dense magnetized winds, and powerful eruptive events. Space-based observatories have revealed that these stars routinely produce superflares with energies of $10^{33}$–$10^{35}$ erg, orders of magnitude more energetic than the largest solar flares on record[1-3]. These flares are frequently accompanied by coronal mass ejections (CMEs), vast magnetized plasma clouds hurled into interplanetary space at speeds exceeding 1,500 km/s. As these fast CMEs propagate outward, they drive interplanetary shocks and accelerate energetic protons and heavy ions up to 40 GeV into Stellar Energetic Particle (StEP) events[4]. These particles penetrate circumstellar disk environments, ionizing molecular hydrogen, driving non-equilibrium gas-phase chemistry in protoplanetary disks, and potentially stripping the atmospheres of young exoplanets (see Fig. 1)[1,5-9]. Furthermore, the mass and angular momentum carried away by CMEs and winds shape stellar rotational and luminosity evolution. Crucially, recent empirical breakthroughs from high spatial resolution X-ray observations of open clusters have dramatically revised our understanding of early stellar evolution. Deep Chandra and ROSAT surveys of eight rich clusters spanning ages from 7 to 750 Myr demonstrate a stark mass-dependent decay in stellar high-energy output[10]. While sub-solar mass stars ($<0.9M_{\odot}$) undergo a slow, gradual decay in X-ray luminosity, solar-mass stars (0.9–$1.2M_{\odot}$) experience a far more rapid decay. This is accompanied by an abrupt coronal softening and the disappearance of hot plasma ($kT \geq$ 1-3 keV) by 100 Myr, driven by the structural transition from distributed to interface-type magnetic dynamos.

The ultraviolet spectral range, particularly the FUV (912–1700 Å) and NUV (1700–3200 Å), is the primary diagnostic window for tracking the atmospheric and chemical consequences of this evolution. Chromospheric and transition region emission lines (Lyα, C IV, Si IV, N V, Mg II h&k), continuum irradiance, and time-resolved spectroscopy during flares all require space-based UV access. Among operating facilities, HST's Imaging Spectrograph (STIS) and Cosmic Origins Spectrograph (COS) remain the world's premier UV spectrographs, with no planned replacement in the near future[11]. This reality makes Hubble irreplaceable for stellar space weather science for at least the next 15 years.

## 2. Key Science Questions Requiring Hubble's Unique Capabilities

We identify four deeply interconnected primary science questions (Q1–Q4) that can only be addressed with transformative advantage using Hubble over the coming decade.

**Question 1. What are the observational signatures of energetic CMEs and stellar energetic particles (StEPs) from active young stars? How do we use physics-based modeling to guide multi-observatory, multi-wavelength observational campaigns?** While stellar superflares are routinely detected across the electromagnetic spectrum, their associated CMEs remain exceedingly difficult to observe directly. Unlike the Sun, where coronagraph imagery provides direct CME detection, stellar CMEs must be inferred from indirect signatures: blueshifted emission or absorption features in chromospheric and transition region lines during flares (indicative of upward-moving plasma), X-ray and EUV flux dimming analogous to solar CME-associated dimming, and enhanced absorption in UV resonance lines[12,13]. Hubble's STIS and COS, with their high spectral resolution and FUV sensitivity, are uniquely positioned to detect these signatures. Time-resolved UV spectroscopy during energetic flares on young solar-like stars can reveal Doppler-shifted emission components, characterize mass-loss rates from individual events, and constrain CME mass and energy[13-15]. The statistical characterization of CME rates as a function of stellar age, rotation, and magnetic field complexity is a prerequisite for all downstream habitability assessments.

**Question 2: What are the physical characteristics of stellar flares, CMEs, and StEPs from young solar-like stars, and how do they impact early stellar evolution? What is the role of CMEs in mass loss and angular momentum evolution relative to steady-state stellar winds?** The magnetic braking of young stellar rotation is one of the most important and least understood processes in stellar physics. The canonical picture attributes angular momentum loss primarily to magnetized stellar winds, but CME-driven mass loss and the associated removal of magnetic helicity represent potentially comparable channels that have been largely neglected. A single energetic CME can expel up to $10^{21}$ g of mass comparable to many days of steady wind output[15]. If superflare-associated CMEs occur at the observed flare rates (hundreds per year), the integrated mass and angular momentum loss may be substantial. Hubble observations addressing Question 2 require both spectroscopic monitoring of UV emission line asymmetries across a well-characterized sample of young solar-like stars spanning ages 10–600 Myr, rotation periods, and spectral subtypes, as well as precise FUV continuum flux measurements to constrain total radiative energy budgets of individual flare-CME events. When combined with simultaneous X-ray observations (Chandra, XMM-Newton) and optical/infrared photometry (TESS, ground-based), these data enable energy-partition analyses.

**Question 3. How can we design a data-constrained, physics-based stellar space weather modeling suite to properly simulate stellar coronae, winds, CMEs, and StEPs for young solar-like stars?** Sophisticated physics-based MHD models including the Alfvén Wave Solar Model (AWSoM) coupled with particle transport frameworks can successfully reproduce solar CME propagation and SEP acceleration from remote-sensing inputs[16]. Extending these tools to stellar environments requires empirical inputs that only Hubble can provide: absolute FUV flux calibrations, transition region and chromospheric emission measures, and time-resolved spectral diagnostics during eruptive events. Hubble's role in this science question is fundamentally that of a calibration and ground-truth anchor. High-quality UV spectra of flares and quiescent phases establish the

boundary conditions and observational tests for stellar space weather models. These models can then be run forward to predict CME and StEP fluxes at the distances of exoplanets around young host stars and their interaction with planetary magnetospheres[17] directly informing habitability assessments.

**Question 4. How do stellar coronae, winds, superflares, CMEs, and StEPs from young solar-like stars impact the chemistry and retention of protoplanetary disk and exoplanetary atmospheres? What observational evidence of atmospheric erosion or modification can Hubble detect in transmission and emission spectroscopy?** The habitability of an exoplanet depends critically on the radiation and particle environment provided by its host star, particularly during the first billion years of the system's history. FUV and EUV irradiation drives photoionization and photodissociation of atmospheric molecules; CME impacts can erode atmospheres through sputtering and charge exchange; and StEPs can catalyze nitrogen chemistry analogous to the terrestrial NOx production that modulates Earth's ozone layer[1,5-7,18]. Young solar-like stars present the most extreme space weather environments precisely at the epoch when planetary atmospheres are being established. Hubble can address this question through two observational modes: (1) transit spectroscopy of young, close-in planets orbiting active stars, searching for signs of enhanced atmospheric escape (Lyα transit absorption, metastable helium absorption at 10830 Å), and (2) direct observation of FUV/UV flare irradiance profiles to construct the high-energy spectral energy distributions (SEDs) needed as inputs for photochemical atmosphere models[19]. These SEDs are critical inputs for JWST atmospheric characterization observations, making Hubble a necessary precursor and complement to JWST's exoplanet science.

## 3. Cross-Cutting Impacts: X-ray Cluster Trends, Roman Synoptic Synergy, and Exoplanetary Transits

The integration of recent X-ray open cluster data introduces a profound paradigm shift in how we interpret space weather impacts. The empirical discovery that solar-mass stars undergo rapid X-ray dimming and a substantial loss of hot coronal plasma component by 100 Myr implies that the ionizing environment around young solar-like stars is significantly more benign than predicted by traditional activity-rotation models[20]. Widely utilized semi-empirical models predict a prolonged high-activity plateau spanning up to 300 Myr, maintaining elevated flare and ionizing flux baselines[21]. The new cluster data demonstrate an order-of-magnitude (0.5 dex) deficit in high-energy emission over the 100–750 Myr epoch compared to these models. Consequently, photoevaporative atmospheric mass loss and water photolysis rates are significantly lower for planets orbiting young solar-like stars, extending the critical window for volatile retention and surface water preservation into mid-Gyr epochs.

This framework requires immediate cross-calibration with the optical/NIR flare distributions and spot modulation baselines that will be uniquely captured by the Nancy Grace Roman Space Telescope and is well suited to perform time domain surveys of thousands of young stars across distant open clusters[10]. These surveys will use high-precision, high-cadence, large-area synoptic infrared and optical monitoring to map starspot filling factors and flare frequency distributions (FFDs). By establishing a direct empirical bridge between the spot area decay, the X-ray luminosity decay, and the UV line irradiances, joint Hubble, Chandra, and Roman observations will break long-standing degeneracies in magnetic dynamo models (see Table 1).

Furthermore, these stellar space weather trends have immediate, transformative consequences for exoplanetary transit observations. During a transit, the intense FUV/NUV flare irradiance from an active host star triggers rapid photoionization and hydrodynamic escape of the planet's upper atmosphere, which can be dynamically monitored via time-resolved transmission spectroscopy. Hubble observations of Lyα and metastable helium transits provide direct, empirical constraints on these atmospheric mass-loss rates. Crucially, because solar-mass stars experience a rapid drop in coronal temperature and flare activity by 100 Myr, mini-Neptunes and super-Earths orbiting these hosts are spared from total atmospheric stripping, allowing them to retain substantial volatile-rich primordial envelopes. Conversely, young K and early M dwarfs maintain elevated, hard X-ray emission and violent flaring regimes over significantly longer timescales, driving severe atmospheric erosion and water depletion. Hubble's unique capability to track UV transit variations during active flare states provides the necessary baseline to model atmospheric photochemistry, breaking degeneracies in molecular retrievals obtained by JWST and safeguarding future direct-imaging searches for biosignatures.

| Capability / Requirement | Science Justification & Multi-Mission Synergy |
|---|---|
| STIS/COS FUV Spectroscopy (1150–2000 Å) | Access to Lyα (1216 Å), C IV (1548/1551 Å), Si IV (1393/1403 Å), and N V (1239/1243 Å) diagnostics of chromospheric/transition region conditions and CME ejecta. No other operating facility accesses these wavelengths. |
| TIMETAG Mode Spectroscopy | Sub-minute time resolution during flares is essential to resolve flare rise and decay, isolate CME Doppler signatures, and characterize energy release timescales. STIS/COS TIMETAG at R ~ 45,000 is optimal. |
| Simultaneous Multi-Wavelength Triggering | Coordinated ToO programs with Chandra/XMM (X-ray flare tracking) and Roman/TESS (optical/NIR flare frequency). Hubble schedulability for rapid ToO response (<12–24 hr) is a critical operational requirement. |
| NUV Capability (STIS/MAMA, ACS/SBC) | Mg II h&k (2796/2803 Å) and NUV continuum fluxes constrain chromospheric heating; ACS/SBC provides direct FUV photometric monitoring to map radiative energy budgets. |
| Absolute Flux Calibration (≤1%) | Stellar irradiance SEDs used in planetary atmosphere models require absolute UV calibration at the 1% level to propagate meaningful constraints on photodissociation rates and escape models. |
| Transit Spectroscopy (STIS/G140M, G230L) | Lyα and NUV transmission spectroscopy of close-in planets around cluster members to probe hydrodynamic atmospheric escape in response to the mass-stratified stellar UV/EUV environment. |

**Table 1: Summary of required Hubble instrumental capabilities, operational parameters, and cross-mission synergies necessary to resolve early space weather evolution and exoplanetary transit impacts.**

## 4. Synergy with Other Missions and the Path to HWO

### *4.1. Multi-Wavelength Campaigns*

Stellar space weather science is inherently multi-wavelength and multi-facility. Hubble's UV observations gain their full scientific power only when combined with contemporaneous data in complementary bands. X-ray observatories (Chandra, XMM-Newton) constrain the coronal temperature, emission measure, and flare peak energetics, providing real-time triggers for Hubble ToO scheduling. In the optical and near-infrared, TESS and ground-based photometry provide bolometric flare energetics, starspot modulation, and rotation periods. The capabilities of Roman will expand this baseline exponentially, providing large-scale, young cluster synoptic surveys to contextualize Hubble's spectroscopic snapshots. At radio wavelengths (VLA, ngVLA, LOFAR, ATCA), coherent radio bursts serve as the most direct proxy for CME-associated shock acceleration and coronal magnetic field evolution. Finally, JWST provides complementary transit spectroscopy in the infrared to trace molecular atmospheric signatures, relying directly on Hubble’s UV irradiance profiles to interpret photodissociation and chemical networks.

### *4.2. Supporting the Habitable Worlds Observatory (HWO)*

The HWO, recommended by the Astro2020 Decadal Survey as the next flagship space telescope, will utilize direct imaging for spectroscopic characterization of rocky planets in the habitable zones of nearby Sun-like stars. A central prerequisite for interpreting HWO spectra, particularly for identifying true biosignatures versus abiotic false positives, is a detailed understanding of the evolving UV activity environment of the target host stars. Hubble is uniquely positioned to deliver this preparatory science over the next decade. First, Hubble can execute targeted stellar UV activity surveys to characterize the FUV/NUV spectral variability and flare rates of the ~100 highest-priority HWO target stars, building an empirical database of host environments. Second, Lyα transit observations with Hubble establish the baseline for atmospheric mass-loss benchmarks, a technique that will be directly inherited and utilized by HWO. Third, COS and STIS spectrophotometric calibration programs provide the absolute UV flux standards that will anchor HWO's UV channel calibrations. Finally, the empirically derived CME and flare statistics as a function of stellar mass and age are vital for HWO mission design, particularly for assessing planet-atmosphere erosion models. In this sense, Hubble and HWO are not sequential but profoundly synergistic; failing to execute this Hubble program would leave HWO critically under-prepared for its core mission.

## 5. Proposed Major Initiatives and Large-Scale Observing Programs

***5.1. Initiative 1: The Young Star UV Time-Domain Treasury (YSUV-T3)***. We propose a multi-cycle Treasury program targeting a carefully selected sample of ~50 young solar-like stars spanning ages 10–600 Myr in well-characterized young clusters and moving groups (including the TW Hya Association, β Pic Moving Group, Tucana-Horologium, AB Dor, DS TucA, EK Dra, $k^1$Cet, Pleiades, and Hyades). The program will execute high-cadence STIS TIMETAG UV spectroscopy across multiple visits per target to capture flare-resolved spectra and build CME candidate samples via Doppler line asymmetry analysis. It will utilize coordinated ToO triggers from X-ray and optical monitoring, enabling rapid Hubble follow-up of superflares within 12 hours of detection. Quiescent baseline UV SED measurements will quantify intrinsic chromospheric and transition region emission as a function of age and rotation, while time-baseline separations of 1–3 years will sample magnetic activity cycle-induced variations. YSUV-T3 will deliver the definitive empirical database of young solar analog UV environments, anchoring stellar space weather models and directly resolving the mass-dependent stellar activity tracks. Estimated program size: ~400 orbits over 3 cycles.

***5.2. Initiative 2: The HWO Target Star UV Preparatory Survey (HWO-UVPS)***. We propose a systematic spectroscopic survey of the ~100 highest-priority HWO direct-imaging target stars (primarily FGK dwarfs within 25 pc) using STIS and COS. The survey will obtain absolute FUV/NUV flux calibrations and line flux measurements for all major chromospheric and transition region diagnostics, establishing the quiescent UV irradiance spectra needed for HWO atmospheric modeling inputs. It will also perform multi-epoch monitoring (3–5 visits per target over 2–3 years) to characterize UV variability amplitudes, flare rates, and duty cycles for each potential HWO host. HWO-UVPS directly addresses the preparatory science mandate highlighted in Astro2020, ensuring that when HWO begins science operations, the high-energy characterization of its target stars is fully complete. Estimated program size: ~250 orbits over 2 cycles.

## 6. Summary and Conclusions

Young, low-mass solar analogs are the Rosetta stones of stellar and planetary evolution. Characterizing their space weather ecosystem is one of the most pressing challenges in modern astrophysics, with direct, transformative implications for exoplanetary habitability and the core scientific goals of the HWO. The HST, through the unique capabilities of STIS and COS, remains the only operational facility capable of directly characterizing these vital phenomena in the FUV/NUV wavelength range. To maximize the return of Hubble's remaining operational lifetime, we advocate for: (1) dedicated ToO scheduling resources for multi-observatory stellar space weather campaigns, minimizing turnaround times to under 12 hours; (2) the execution of the Young Star UV Time-Domain Treasury (YSUV-T3) program to chart mass-stratified flare-CME evolution; (3) the approval of the HWO Target Star UV Preparatory Survey (HWO-UVPS) as an explicitly protected preparatory initiative; and (4) enhanced coordination protocols between Hubble, Chandra, and Roman to maximize simultaneous multi-wavelength flare observations. Hubble's legacy in the 2030s will be measured not only by its standalone discoveries, but by the empirical foundation it lays for the next generation of flagships.